\documentclass[preprintnumbers,nofootinbib,noshowpacs,prd,superscriptaddress,longbibliography]{revtex4}

\preprint{PITT-PACC 1406}

\usepackage{graphicx}
\usepackage{amsmath,amssymb}
\usepackage{url}
\usepackage{color}
\usepackage{multirow}
\usepackage{subfigure}
\usepackage{verbatim}
\usepackage{comment}
\usepackage{hyperref}
\usepackage{axodraw4j}
\usepackage{pstricks}

\newcommand{\lsim}{\mathrel{\mathop{\kern 0pt \rlap
  {\raise.2ex\hbox{$<$}}}
  \lower.9ex\hbox{\kern-.190em $\sim$}}}
\newcommand{\gsim}{\mathrel{\mathop{\kern 0pt \rlap
  {\raise.2ex\hbox{$>$}}}
  \lower.9ex\hbox{\kern-.190em $\sim$}}}
\newcommand{\gev}{{\,{\rm GeV}}}
\newcommand{\tev}{{\,{\rm TeV}}}
\newcommand{\beq}{\begin{equation}}
\newcommand{\eeq}{\end{equation}}
\newcommand{\bea}{\begin{eqnarray}}
\newcommand{\eea}{\end{eqnarray}}

\def\mm{\mu^+ \mu^-}

\def\fbi{{\rm fb}^{-1}}
\def\abi{{\rm ab}^{-1}}

\def\cm2s{{\rm cm^{-2} s^{-1}}}

\begin{document}

\title{Radiative Return for Heavy Higgs Boson at a Muon Collider}
\bigskip
\author{Nabarun Chakrabarty}
\email{nabarunc@hri.res.in}
\affiliation{Regional Centre for Accelerator-based Particle Physics, Harish-Chandra Research Institute, Chhatnag Road, Jhusi, Allahabad - 211 019, India}
\author{Tao Han}
\email{than@pitt.edu}
\affiliation{Pittsburgh Particle physics, Astrophysics and Cosmology Center, Department of Physics $\&$ Astronomy,
University of Pittsburgh, 3941 O'Hara St., Pittsburgh, PA 15260, USA}
\affiliation{Center for High Energy Physics, Department of Physics, Tsinghua University, Beijing 100084, P.R. China}
\affiliation{Korea Institute for Advanced Study (KIAS), Seoul 130-012, Korea}
\author{Zhen Liu}
\email{zhl61@pitt.edu}
\affiliation{Pittsburgh Particle physics, Astrophysics and Cosmology Center, Department of Physics $\&$ Astronomy,
University of Pittsburgh, 3941 O'Hara St., Pittsburgh, PA 15260, USA}
\author{Biswarup Mukhopadhyaya}
\email{biswarup@hri.res.in}
\affiliation{Regional Centre for Accelerator-based Particle Physics, Harish-Chandra Research Institute, Chhatnag Road, Jhusi, Allahabad - 211 019, India}

%
\begin{abstract}
Higgs boson properties could be studied with a high accuracy at a muon collider via the $s$-channel resonant production. We consider the situation where the center-of-mass energy of the muon collider is off the resonance above the Higgs mass. We discuss the discovery potential for a generic heavy Higgs boson ($H$) and compare different production mechanisms, including  the ``radiative return'' ($\gamma H$), $Z$-boson associated production ($ZH$) and heavy Higgs pair production ($HA$).
These production mechanisms do not sensitively rely on {\it a priori} knowledge of the heavy Higgs boson mass.  
We include various types of Two Higgs Doublet Models for the comparison. We conclude that  the radiative return process could provide an important option for both the heavy Higgs discovery and direct measurement of invisible decays at a high energy muon collider.
\end{abstract}
\pacs{}
\maketitle

\section{Introduction}

With the discovery of the Standard Model (SM)-like Higgs boson ($h$) at the Large Hadron Collider (LHC)~\cite{Aad:2012tfa, Chatrchyan:2012ufa}, the follow-up examinations of its properties at the LHC and future colliders will be of high priority for collider physics.
While an electron-position collider near the $Zh$ threshold or utilizing weak boson fusion as a ``Higgs Factory'' may provide high precision measurement for its couplings in a model-independent way~\cite{Peskin:2012we,Asner:2013psa,Han:2013kya}, a muon-antimuon collider could directly and accurately determine its total width, mass and couplings via the $s$-channel resonant production of a Higgs boson \cite{Barger:1995hr,Barger:1996jm,Han:2012rb,Conway:2013lca,Alexahin:2013ojp}.

However, the Higgs sector may not be as simple as it is in the minimal electroweak theory. A wide class of new physics scenarios, ranging from supersymmetry (SUSY) \cite{Gunion:1984yn} to models of neutrino mass generation \cite{Konetschny:1977bn,Cheng:1980qt,Lazarides:1980nt,Schechter:1980gr,Mohapatra:1980yp}, postulates the existence of an extended sector of fundamental scalars.
While such an extension could leave some imprint on the properties of the recently discovered Higgs boson, it is also imperative that the proposed future colliders should have the potential to identify additional scalars that could be produced within its kinematic reach. Due to the rather weak couplings and the large SM backgrounds, the LHC will have limited coverage for such search~\cite{Gianotti:2002xx,Lewis:2013fua,Li:2013nma,Dawson:2013bba,Coleppa:2014hxa}. At a future lepton collider, on the other hand, due to the clean experimental environment, it would be straightforward to identify a heavy Higgs signal once it is copiously produced on resonance~\cite{Eichten:2013ckl}.

The exact center-of-mass energy required for an optimal heavy Higgs signal depends on the unknown heavy Higgs mass, in particular for the $s$-channel resonant production at a muon collider. The situation may be remedied if instead we consider associated production of a Higgs boson with other particles.
A particularly interesting process is the ``radiative return'' (RR) process. In the case of the Higgs boson production, the processes under consideration are
\bea
\mu^{+} \mu^{-} \rightarrow  \gamma H, \gamma A ,
\label{eq:rr}
\eea  
where $H$ ($A$) is a heavy neutral CP-even (CP-odd) state, respectively. 
When the center of mass energy of the muon collider is above the heavy Higgs resonance, the photon emission from the initial state provides an opportunity of the heavy Higgs boson ``back'' to the resonance.
For this, one does not need to know the mass of the (unknown) heavy scalar.
This mechanism alone could also provide an excellent channel to measure the invisible decay of the heavy Higgs
boson. 
Without losing generality, we illustrate our main points with a notation in the context of a two-Higgs-doublet model 
(2HDM) \cite{Haber:1978jt}, where the vacuum expectation values (vev) of both the doublets contribute to the $W$- and $Z$-masses.  

In Sec.~\ref{sec:RR}, 
we first present the radiative return production of heavy Higgs boson in $\mu^{+} \mu^{-}$ collision in detail. We also consider the production $l^{+} l^{-} \rightarrow ZH$ and $l^{+} l^{-} \rightarrow AH\ (l=e,\mu)$ in Sec.~\ref{sec:pair}.
To make the illustration more concrete, we compare these production modes in Sec.~\ref{sec:comp} in the framework of 2HDM. Because of the rather clean experimental environment and the model-independent reconstruction of the Higgs signal events at lepton colliders, we also study the sensitivity of the invisible decay from
the radiative return process in Sec.~\ref{sec:invi}. Finally, we summarize our results and conclude in Sec.~\ref{sec:sum}.

\section{Production Mechanisms}
\label{sec:prod}

\begin{figure}[t]
\centering
\subfigure[ $~H/A$ \lq\lq{}Radiative Return\rq\rq{}]{\includegraphics[scale=0.5]{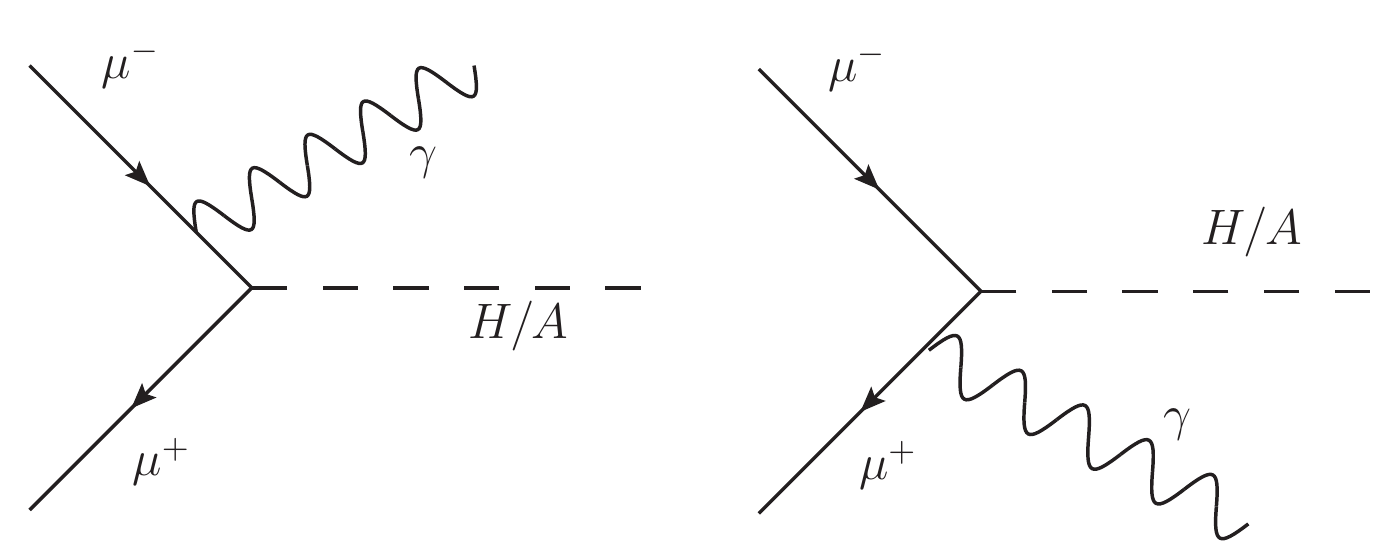}}
\label{fig:Feyn_RR}
\subfigure[ $~ZH$ associated production]{\includegraphics[scale=0.5]{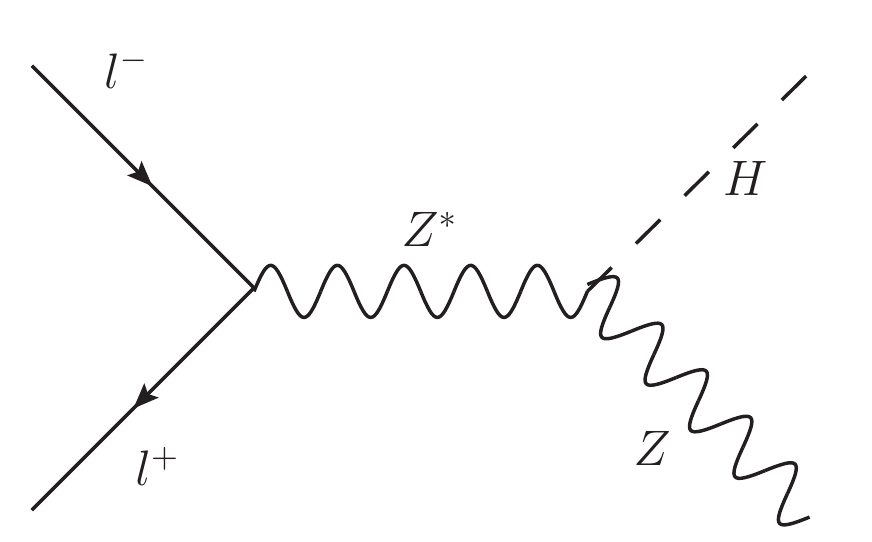}}
\label{fig:Feyn_RR}
\subfigure[ $~HA$ pair production]{\includegraphics[scale=0.5]{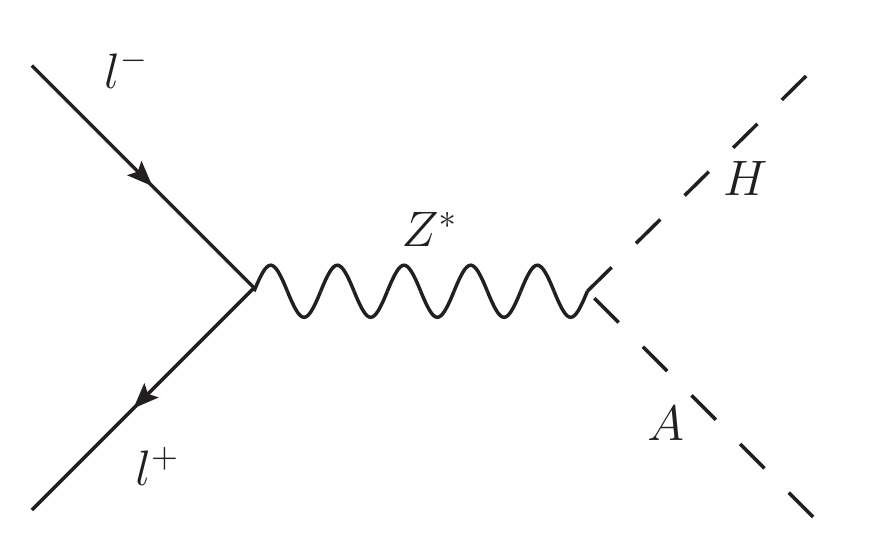}}
\label{fig:Feyn_RR}
\caption{Main production mechanisms of heavy Higgs boson $H/A$ at lepton colliders.}
\label{fig:Feyn}
\end{figure}

Perhaps the most useful feature of a muon collider is the potential
to have $s$-channel resonant production of the Higgs boson~\cite{Barger:1995hr,Barger:1996jm,Han:2012rb,Eichten:2013ckl,Alexahin:2013ojp}. As has been already mentioned in the previous section, such a machine 
undoubtedly has its merits in analyzing in detail the already discovered Higgs boson near 125 GeV. When it comes to identifying a heavier additional (pseudo)scalar, however, we do not have any {\it a priori} knowledge about the mass, rendering the new particle search rather difficult. If one envisions a rather wide-ranged scanning, it would require to devote a large portion of the design integrated luminosity~\cite{Conway:2013lca,Alexahin:2013ojp}. In this section,
we discuss the three different production mechanisms for the associated production of the heavy Higgs boson. Besides the ``radiative return'' as in Eq.~(\ref{eq:rr}), we also consider 
\bea
\mm\to Z^*\to ZH\ {\rm and}\ HA.
\label{eq:zh}
\eea
The relevant Feynman diagrams are all shown in Fig.~\ref{fig:Feyn}.

We first parametrize the relevant heavy Higgs boson couplings as
\bea
\label{eq:int}
\mathcal{L}_{int} &=& -\kappa_\mu \frac {m_\mu} {v} H \bar \mu \mu + i \kappa_\mu \frac {m_\mu} {v} A \bar \mu \gamma_5 \mu
    + \kappa_Z \frac {m_Z^2} {v} H Z^\mu Z_\mu 
    +\frac {g} {2\cos\theta_W} \sqrt{ (1-\kappa_Z^2)} (H\partial^\mu A-A\partial^\mu H) Z_\mu.
\eea

\begin{table}[t]
\centering
\begin{tabular}{|c|c|c|c|}
  \hline
  Coupling & $\kappa\equiv g/g_{\rm SM}$ & Type-II \& lepton-specific& Type-I \& flipped\\ \hline
  $g_{H\mu^+\mu^-}$ & $\kappa_\mu$ &$\sin\alpha/\cos\beta$ & $\cos\alpha/\sin\beta$ \\ \hline
  $g_{A\mu^+\mu^-}$ & $\kappa_\mu$ &$\tan\beta$ & $-\cot\beta$ \\ \hline
  $g_{HZZ}$ & $\kappa_Z$ & $\cos(\beta-\alpha)$ & $\cos(\beta-\alpha)$ \\ \hline
  $g_{HAZ}$ & $1-\kappa^2_Z$ & $\sin(\beta-\alpha)$ & $\sin(\beta-\alpha)$ \\
  \hline
\end{tabular}
\caption[]{Parametrization and their 2HDM models correspondence.}
\label{tab:parameters}
\end{table}

The two parameters $\kappa_\mu$ and $\kappa_Z$ characterize the coupling strength with respect to the SM Higgs boson couplings to $\mu^+\mu^-$ and $ZZ$. The coupling $\kappa_\mu$ controls the heavy Higgs resonant production and the radiative return cross sections, while $\kappa_Z$ controls the cross sections for $ZH$ associated production and heavy Higgs pair $HA$ production.
We have used $\kappa_\mu$ as the common scale parameter for Yukawa couplings of both the CP-even $H$ and the CP-odd $A$, although in principle they could be different. For the $HAZ$ coupling we have used the generic 2HDM relation: $\kappa_Z$ is proportional to $\cos(\beta-\alpha)$ and the $HAZ$ coupling is proportional to 
$\sin(\beta-\alpha)$.\footnote{Customarily,  $\tan\beta$ is the ratio of the two vev's, and $\alpha$ is the mixing angle of the two scalar states.} In the heavy Higgs decoupling limit of 2HDM at large $m_A$, $\kappa_{Z}\equiv \cos(\beta-\alpha)\sim m_{Z}^{2}/m_{A}^{2}$ is highly suppressed and $\kappa_\mu \approx \tan\beta\ (-\cot\beta)$ in Type-II \cite{Hall:1981bc,Donoghue:1978cj} and lepton-specific \cite{Barger:1989fj,Aoki:2009ha,Pich:2009sp,Branco:2011iw} (Type-I \cite{Haber:1978jt,Hall:1981bc} and flipped \cite{Barger:1989fj,Aoki:2009ha,Pich:2009sp,Branco:2011iw}) 2HDM.
Note that many SUSY models, including MSSM and NMSSM, are essentially Type-II 2HDM, subject to fewer tree-level parameters for the Higgs potential and potentially large supersymmetric loop corrections. We tabulate our choices of parameters and their 2HDM correspondences in Table.~\ref{tab:parameters}. 
We reiterate that such a notation can be carried over to any scenario
where there is another multiplet in addition to the SM Higgs doublet
contributing to the $W$- and $Z$-masses, whereby the $WW$ and $ZZ$ couplings of
the two neutral CP-even scalars are connected by a unitary relationship,
with some SU(2) Clebsch-Gordan coefficients arising in addition.

\begin{table}[t]
\centering
\begin{tabular}{|c|c|}
  \hline
  $\sqrt{s}=1.5~\gev$ & $500~\fbi$ \\ \hline
  $\sqrt{s}=3.0~\gev$ & $1,760~\fbi$ \\  \hline  
 \multicolumn{1}{|r}{Beam energy spread:} & $R=0.1\%$ \\ \hline \hline
 \multicolumn{1}{|r}{Polar angle acceptance:} & $10^\circ<\theta<170^\circ$  \\ \hline
 \multicolumn{1}{|r}{$p_{T \rm min}$ for photon:} & $10~\gev$ \\ \hline
 \multicolumn{1}{|r}{Photon Energy Resolution:} & $0.17/\sqrt{E}\oplus 0.01$ \\ \hline
 \multicolumn{1}{|r}{$p_{T \rm min}$ for lepton:} & $20~\gev$ \\ \hline
 \multicolumn{1}{|r}{$\Delta R_{\rm min}$ for leptons:} & $0.2$ \\ \hline
\end{tabular}
\caption[]{Muon Collider Parameters~\cite{Delahaye:2013jla} assuming four collider years of running. The photon energy resolution is set as SiD from ILC TDR~\cite{Behnke:2013lya}.}
\label{tab:MuC}
\end{table}

\begin{figure}[t]
\centering
\includegraphics[scale=0.36]{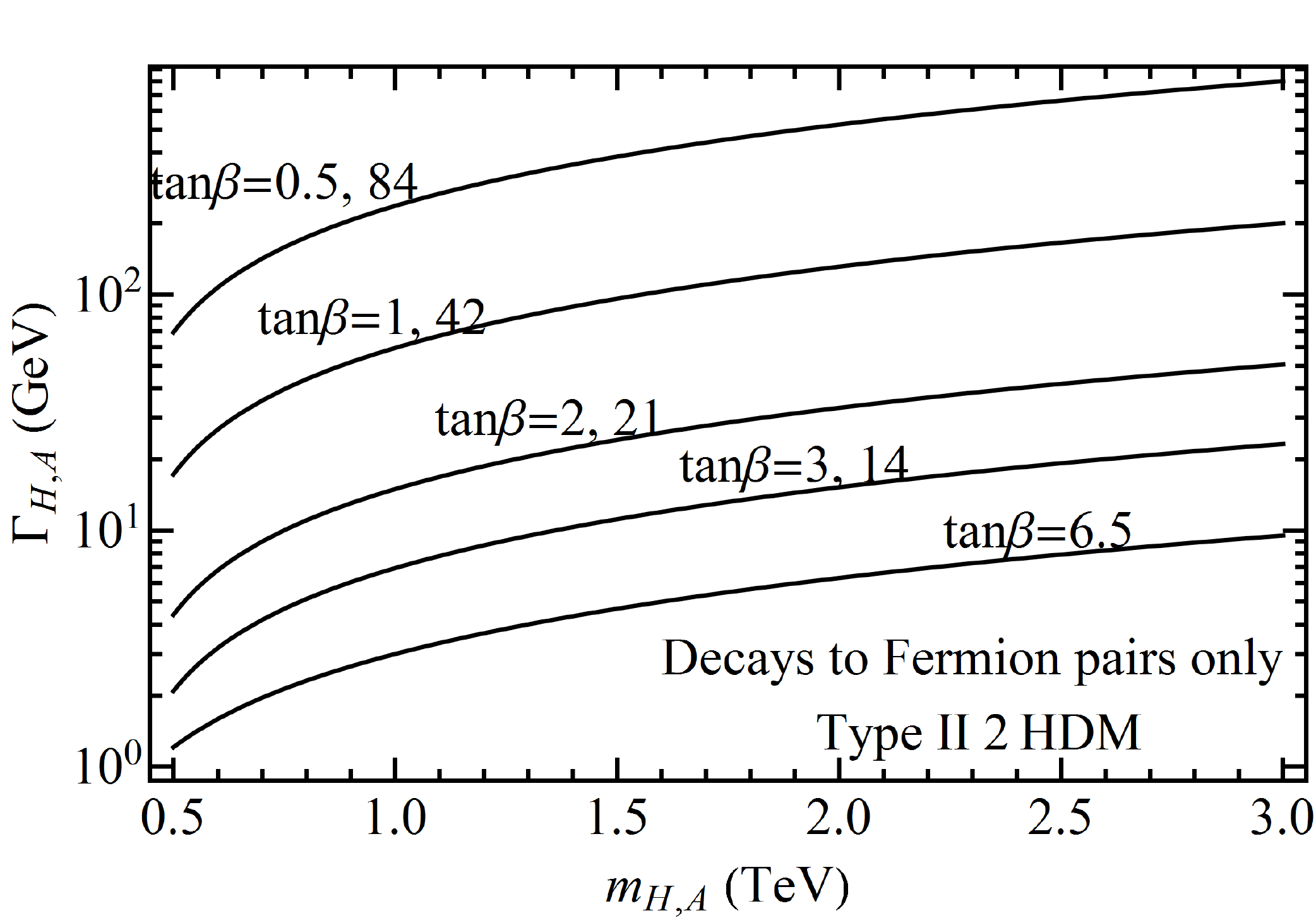}
\caption{Total width of heavy Higgs boson in Type II 2HDM as a function of Higgs mass for a variety values of $\tan\beta = \kappa_\mu$. We only consider partial widths to fermion pairs here. The total width is symmetric with respect to $\sqrt{m_t/m_b}$.}
\label{fig:width}
\end{figure}

We choose the following configuration as shown in Table~\ref{tab:MuC}
for the muon collider parameters and the detector acceptance, to study feasibilities of these different
production channels. The beam energy spread is defined as

\begin{equation}
\frac {d \mathcal{L} (\sqrt{s})} {d\sqrt {\hat{s}}} = \frac {1} {2\pi\Delta} {\rm exp}[-\frac {(\sqrt{s}-\sqrt{\hat{s}})^2} {2\Delta^2}],
\end{equation}

with $\Delta=R\sqrt{s}/\sqrt{2}$.


\subsection{Radiative Return}
\label{sec:RR}

Due to the ``radiative return'',
when the heavy Higgs boson mass is below the center of mass energy of the muon collider, the photon emission from the initial state provides an opportunity of the heavy Higgs boson ``back'' to resonance. The signature is quite striking: a mono-chromatic photon plus other recoil particles. The ``recoil mass'' is a sharp resonant peak at $m_{H/A}^{}$, manifesting itself from the continuous background. This photon's energy is subject to the beam
energy spread and detector energy smearing. The tagging of the heavy Higgs boson from
its decay product, if necessary, provides extra handle on reducing the background and increasing the significance. 

\subsubsection{Signal and Background}

The characteristics of this RR signal is a photon with the energy given by
\beq
E_\gamma= \frac {\hat s - m_{H/A}^2} {2\sqrt{\hat s}},
\eeq
from which one constructs a recoil mass peaked at the heavy Higgs mass $m_{H/A}$. The energy of this photon is smeared by the following factors: detector photon energy resolution, collider beam energy spread, additional (soft) ISR/FSR, and heavy Higgs total width. Our choice of the detector photon energy resolution and beam energy spread are as shown in Table~\ref{tab:MuC}. The beam energy spread and (soft) ISR are of GeV level~\cite{Delahaye:2013jla}.
When the Higgs boson mass is significantly below the beam energy, the recoil mass construction receives large smearing due to the energy resolution for the very energetic photon. 

Besides the Higgs boson mass, the other most important parameter is the total width, which effectively smears the mono-chromatic photons as well. We calculate the total width as a sum of the partial widths to fermion pairs for Type II 2HDM in Fig.~\ref{fig:width}. In this model, $\kappa_\mu=\tan\beta$ in the decoupling limit. The total width is minimized when $\tan\beta=\sqrt{m_t/m_b}$. Because of the quadratic dependence, there are typically two values to give the same width $\tan\beta_{1}\cdot \tan\beta_{2}={m_t/m_b}$. Numerically we take $m_t/m_b = 42$. We can see that typically the total width ranges from a few GeV to hundreds of GeV. The total width of heavy Higgs boson could remain small in lepton-specific 2HDM. We thus choose three representative values for the total width: 1, 10, and 100 GeV for later discussions.

The inclusive cross section for mono-photon background is very large in comparison with the radiative return signal. The background is mainly from the M\"oller scattering with ISR/FSR  $\mm \to \mm \gamma$, 
and the $W$ exchange with ISR $\mm \to \nu\nu \gamma$. 
The signal background ratio is typically of the order $10^{-3}$ for a 3 TeV muon collider. As a result, for the discovery through the RR process, we need to rely on some exclusive processes, or to the least veto mono-photon plus missing energy and mono-photon plus dimuon exclusive channels.

\begin{figure}[t]
\centering
\includegraphics[scale=0.36]{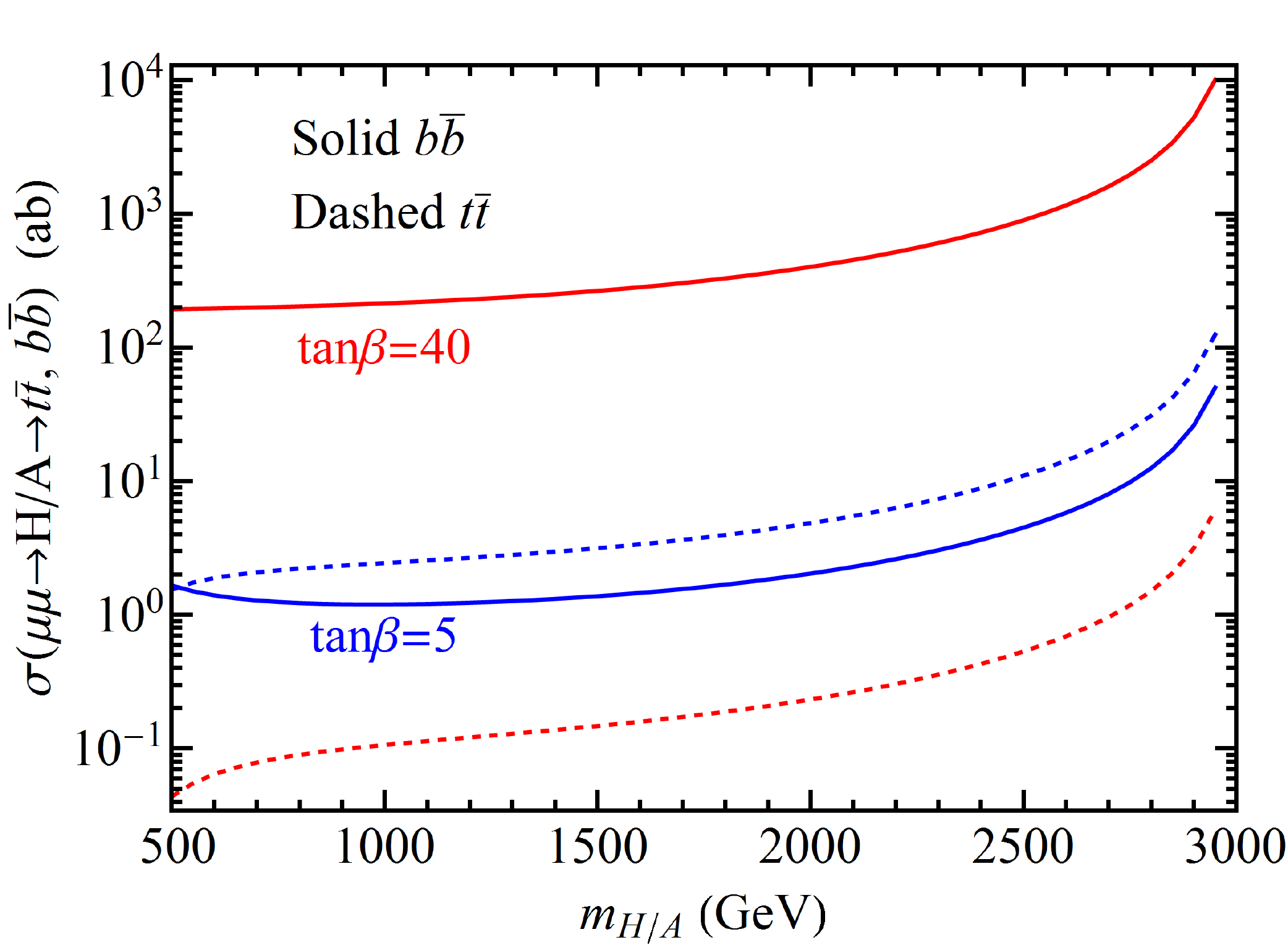}
\includegraphics[bb=0.0in -0.1in 7in 6.0in,width=212pt]{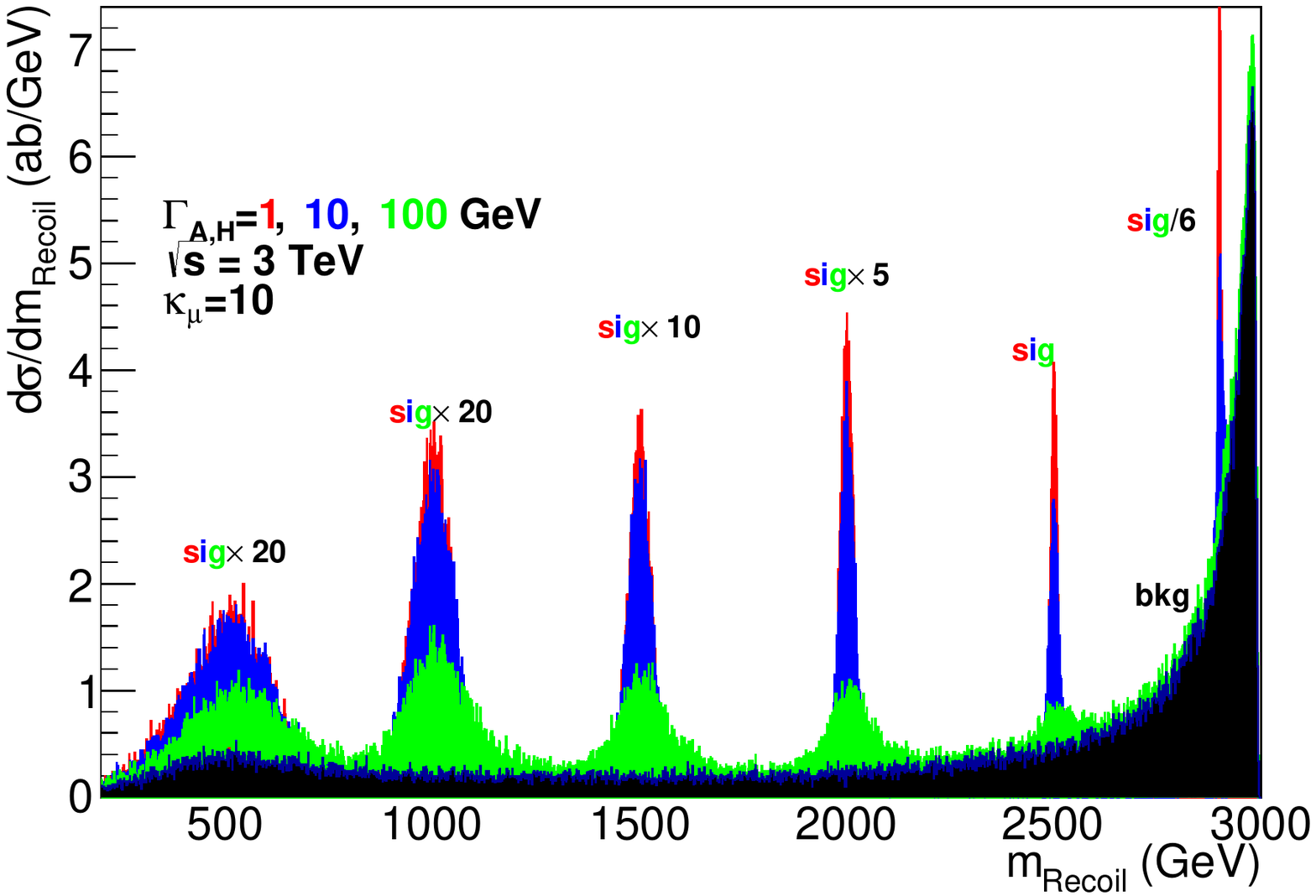}
\caption{Left panel: total cross section for $H/A \to b\bar b$ (solid lines) and $t\bar t$ (dashed lines) as a function of $m_{H/A}^{}$ at $\sqrt{s} = 3$ TeV, in Type-II 2HDM scenario for $\tan\beta$ = 5 (blue) and 40 (red). Right panel: recoil mass distribution for heavy Higgs mass of 0.5, 1, 1.5, 2, 2.5, 2.9 TeV with total width 1 (red), 10 (blue), and 100 (green) GeV at a 3 TeV muon collider. The beam energy resolution and photon energy resolution are as shown in Table.~\ref{tab:MuC}. ISR and FSR are included but not beamstrahlung. Background (black) includes all events with a photon that has $p_T>10~\gev$. Note that signal and background have different multiplication factors for clarity.}
\label{fig:sigbkg}
\end{figure}

It should be noted that, in a 2HDM, the heavy neutral scalar $H$ may decay into both $t\bar t$, $b\bar b$ and $\tau^+\tau^-$ modes, where the branching ratios depend on $\tan\beta$. We adopt the Type-II 2HDM for illustration. We show in Fig.~\ref{fig:sigbkg} the total cross sections (left panel) for $\mu^{+} \mu^{-} \rightarrow H/A  \gamma \rightarrow q\bar q \gamma$ (for $q = t,b$) for $\tan\beta=5, 40$, with the basic cuts applied on the photon. It is clear from the plots that while the rates for $t\bar t \gamma$ is considerably suppressed for large values of $\tan\beta$, it can be of comparable magnitude (or even larger) to that for $b\bar{b} \gamma$ for relatively low $\tan\beta$. 
Judicious criteria for event selection, therefore, need to be developed for both channels.
In the rest of the present study, however, only the $b\bar b$ mode is considered for simplicity.

To be more specific, we choose the $b\bar b$ final state as a benchmark with heavy Higgs boson decay branching fraction (Br) to this final state to be $80\%$. We also assume $80\%$ $b$-tagging efficiency and require at least one $b$-jet tagged. In fact, any visible decay of the heavy Higgs boson except for the dimuon final state, negligible in most of models, 
would be very efficient in background suppression. One could also interpret our assumption as that $80\%$ of the decays of the Higgs boson could be utilized.

We employ Madgraph5~\cite{Alwall:2011uj} and for parton level signal and background simulations and tuned Pythia 6.4~\cite{Sjostrand:2006za} mainly for ISR and FSR, and further implement detector smearing and beam energy spread with our own code.
We show the recoil mass distribution for the heavy Higgs boson mass of 0.5, 1, 1.5, 2, 2.5, 2.9 TeV each with 1, 10, 100 GeV width at a 3 TeV muon collider in Fig.~\ref{fig:sigbkg} (right panel). Both cross sections of the signal and the background at fixed beam energy increase as the recoil mass increases due to the infrared nature of the photon radiation.
The spread of recoil mass peak increases at a lower mass, due to the larger photon energy detector resolution smearing at a higher photon energy. We can see that the pronounced mass peaks look promising for the signal observation, and the 
RR process is a plausible discovery production mechanism that does not rely on the precise knowledge of the new heavy Higgs boson mass. We discuss the observability of this mode in next subsection. 

\subsubsection{Estimated Sensitivities}

\begin{figure}[t]
\centering
\subfigure{\includegraphics[scale=0.6]{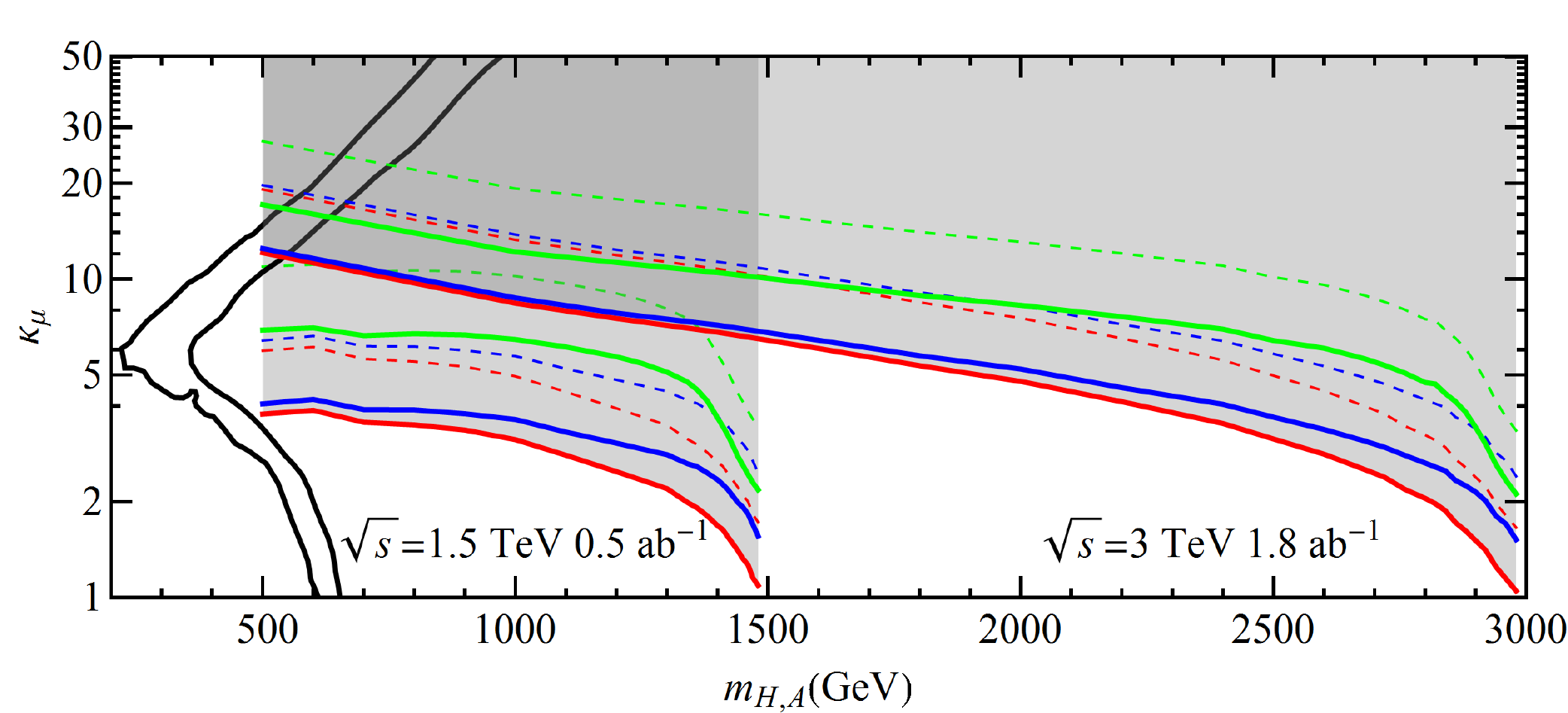}}
\caption[]{Estimated $2\sigma$ exclusion limits (solid) and $5\sigma$ discovery limits (dashed) in the Higgs mass and $\kappa_\mu$ plane, shown as shaded region. We include the cases with Higgs width 1 (red), 10 (blue), and 100 (green) GeV. We overlay the 3 TeV muon collider reach (dark shade) over 1.5 TeV muon collider results (light shade). For comparison, the two solid black wedged curves reproduce the LHC coverage in $m_{A}$\nobreakdash-$\tan\beta$ plane for 300~$\fbi$ and 3000~$\fbi$, respectively.
}
\label{fig:RRsensi}
\end{figure}

To quantify the reach of the signal observation, we choose different bin sizes according to the spread of the photon energy distribution. This is because the recoil mass spread is broader than the photon energy smearing, as scaled by a factor of $\sqrt{\hat s}/m_{H/A}$. This implies the Higgs mass resolution would be much worse than the photon energy resolution if the mass is far away from the beam energy. 
We find the bin sizes in step of 1 GeV that optimize statistical significance of signal at $\kappa_\mu=10$ over the background. With this optimal choice of number of bins, we show the 2$\sigma$ exclusion (solid) and 5$\sigma$ discovery (dashed) limits from RR in Fig.~\ref{fig:RRsensi} for both 1.5 TeV and 3 TeV muon colliders as described in Table~\ref{tab:MuC}, for three different benchmark heavy Higgs width values 1, 10, and 100 GeV in red, blue, and green, respectively. The results show that the RR production mode could cover a large $\kappa_\mu$ ($\tan\beta$ in Type II 2HDM) region. 
To put these results into perspective, we reproduce the LHC curves for the discovery reach on the $m_A-\tan\beta$ plane in solid black lines for 300~$\fbi$ and 3000~$\fbi$~\cite{Gianotti:2002xx}. These LHC discovery projections are mainly from searches on heavier Higgs bosons decaying into SM particles such as $\tau^+\tau^-$ and $t\bar t$, in the maximal mixing scenario in the MSSM. This \lq\lq{}wedge\rq\rq{} shape indicates the LHC\rq{}s limitation in discovering heavy Higgs bosons in the medium $\tan\beta$ range, roughly when the production rate is minimal for the MSSM as  a Type-II 2HDM. 
It is important to see the significant extension at the high energy muon collider via the RR process over the LHC coverage in the heavy Higgs parameter space.

\subsection{$ZH$ Associated Production and $HA$ Pair Production}
\label{sec:pair}

The $ZH$ associated production and $HA$ pair production of Eq.~(\ref{eq:zh})
at tree level are mediated by an off-shell $Z$ boson. The cross section for the $ZH$ associated production is proportional to $\kappa_Z^{2}$. On the other hand, the $HA$ pair production is proportional to $1-\kappa_Z^2$ in generic
2HDM models. These two channels bear some complementarity with each other.
To quantify our study, we assume 90\%
tagging efficiency for the visible $Z$ decays in the $ZH$ associated
production. We also studied the leptonic $Z$ boson decay mode, where requirement on lepton $p_T$, angle and separation are imposed as described in Table~\ref{tab:MuC}. For simplicity, we take both the CP-even and CP-odd heavy
Higgs bosons to have the same mass.

\begin{figure}[t]
\centering
\subfigure{\includegraphics[scale=0.37]{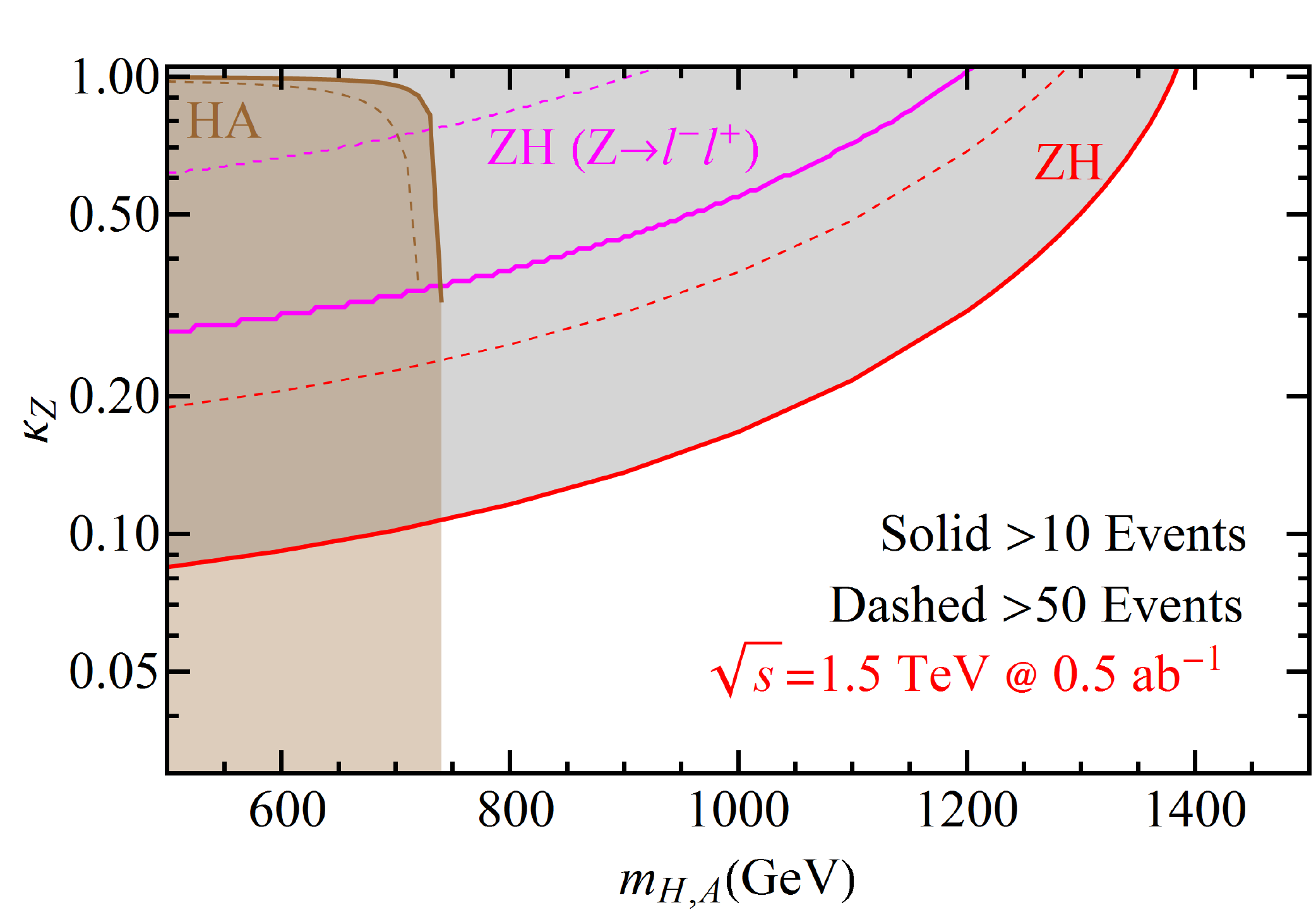}}
\subfigure{\includegraphics[scale=0.38]{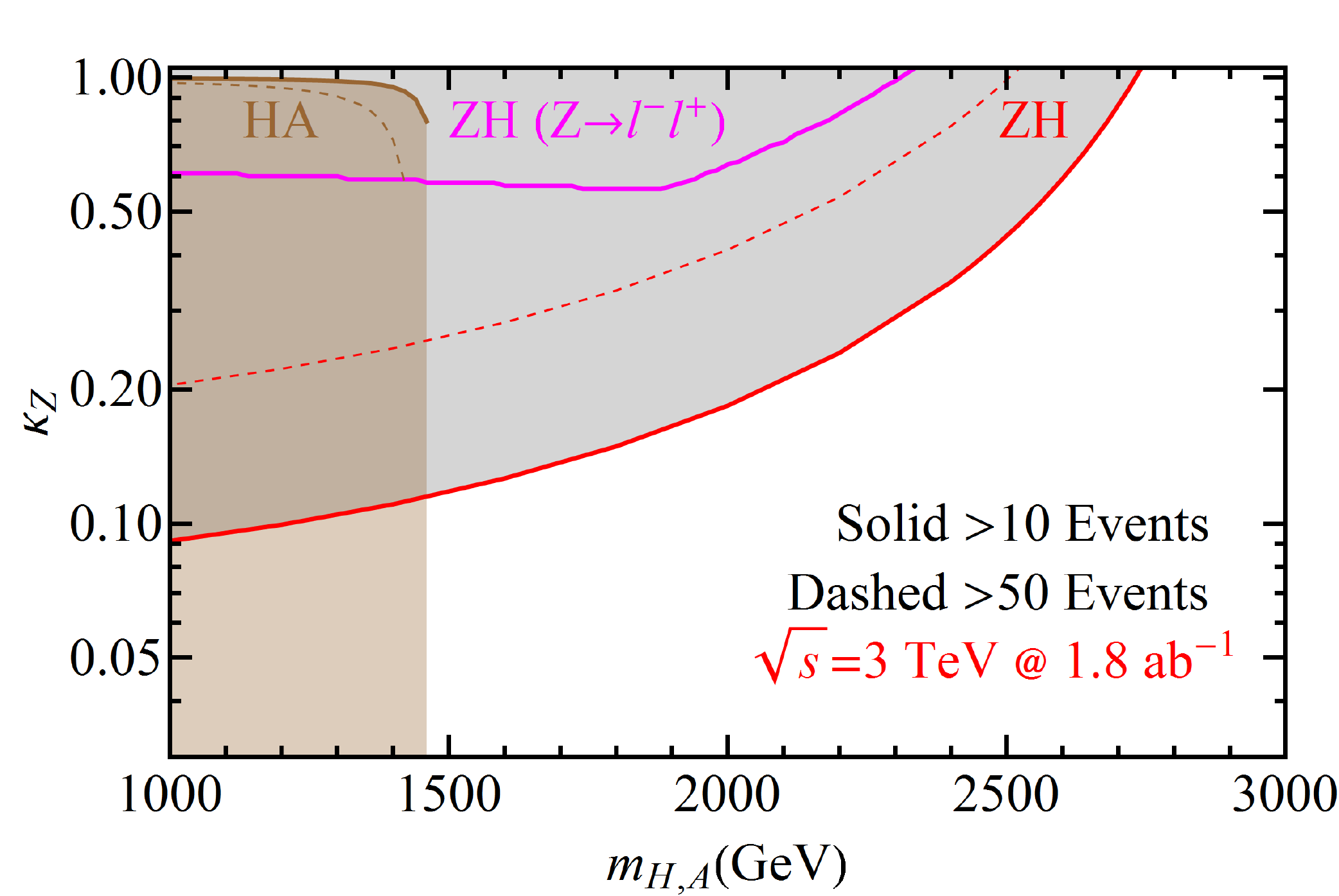}}
\caption{Sensitivity to the Higgs coupling $\kappa_{Z}$ versus the Higgs mass for the $ZH$ associated production (red for all visible $Z$ decays, magenta for the leptonic $Z$ decay only) and $HA$ pair production (brown) 
for the muon collider defined as in Table~\ref{tab:MuC} at the center of mass energy $1.5~\tev$ (left panel) and $3~\tev$ (right panel). Shaded regions bounded  by solid (dashed) curves are regions with more than 10 (50) signal
  events being produced, indicating the exclusion (discovery) reach.}
\label{fig:ZHHA}
\end{figure}

In Fig.~\ref{fig:ZHHA} we show the event contours with 10 events (solid) and 50 events (dashed) 
for both $ZH$ and $HA$ channels in the $m_{H,A}$\nobreakdash-$\kappa_{Z}$ plane. 
As expected, once crossing the kinematical threshold, the $HA$ channel would be sensitive to a large range of $\kappa_{Z}$ value. For instance,  even for $\kappa\sim 0.97$,
one still have $6\%$ of the full cross section which leads to about 15 events.
The kinematically favored channel $ZH$ associated production is more sensitive than the $HA$ pair production, expending to a larger $m_{H}$ region, as long as $\kappa_{Z}>0.1$. 
A higher energy collider would extend the mass coverage to the multiple TeV kinematical limit, with a proportionally larger $\kappa_Z$ value as seen in the figures.

\subsection{Comparison of Different Modes}
\label{sec:comp}

Kinematically, the RR process and the $ZH$ associated production have quite different threshold behavior
due to the massless nature of  the photon. The closer the Higgs boson mass is to the energy threshold, the more effective the RR channel would be with respect to the $ZH$ associated production.
Well above the threshold on the other hand, these two processes scale with the energy in the same way as $1/s$.
Dynamically, the RR process is only dependent on $\kappa_\mu$, while both $ZH$ associated production and $HA$
pair production mainly depend on $\kappa_Z$. These two parameters are essentially independent of each other, characterizing the muon Yukawa coupling and the Higgs-gauge coupling, respectively.

\begin{figure}[t]
\centering
\includegraphics[scale=0.382]{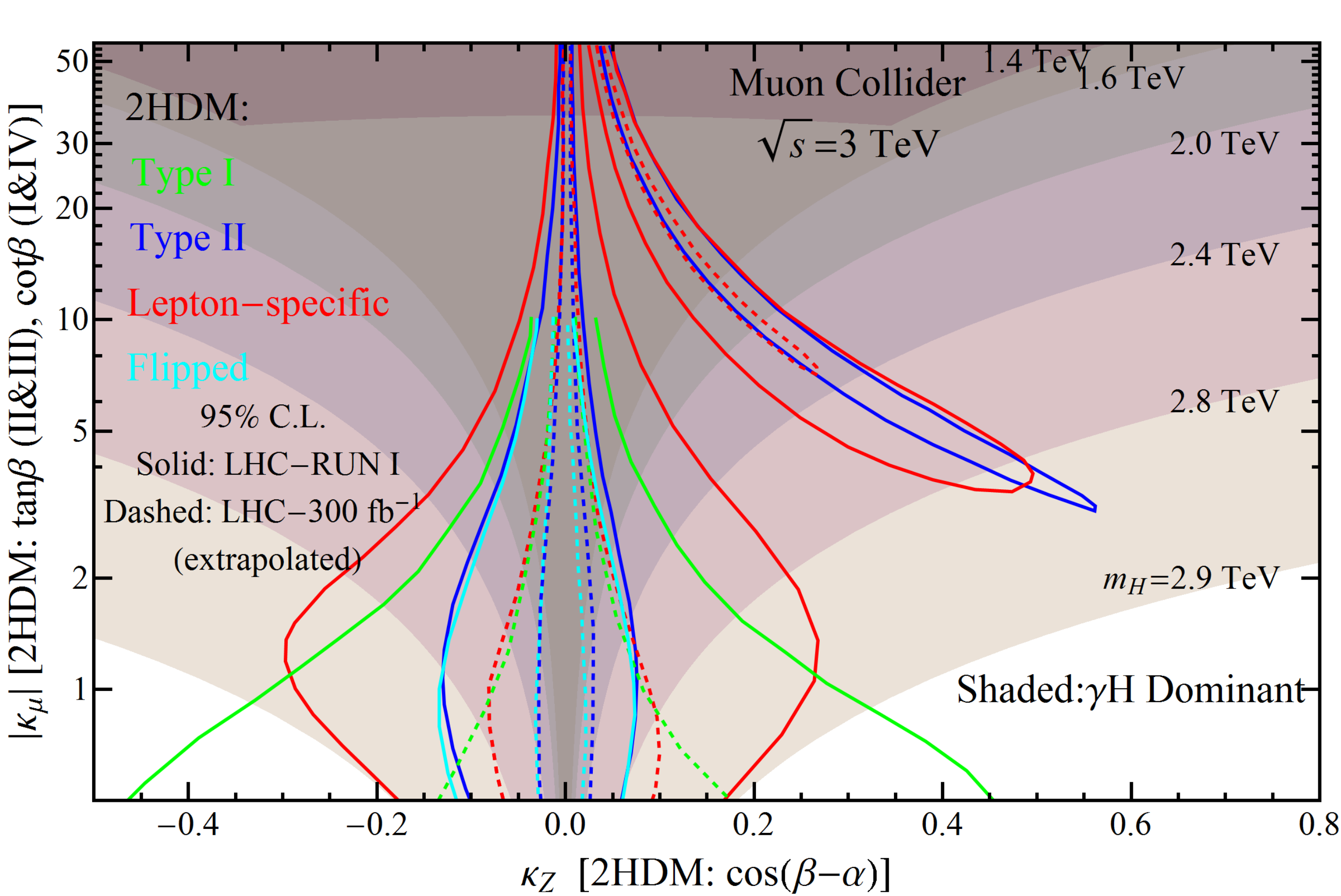}
\caption[5pt]{Comparison of sensitivities between different production mechanisms in the parameter plane $\kappa_\mu$\nobreakdash-$\kappa_Z$ for different masses of the heavy Higgs boson at the $3~\tev$ muon collider. The shaded regions show the higher signal rate from the RR process than both the $ZH$ associated production and $HA$ pair production. 
We also reproduce the allowed parameter regions (extracted from Ref.~\cite{Barger:2013ofa}) for
  four types of 2HDM with current LHC data (solid) and projection  after LHC-$300~\fbi$ (dashed).}
\label{fig:kappas}
\end{figure}

It would be nevertheless informative to put side-by-side the reach of the two theory parameters via these two processes. Our results are summarized in Fig.~\ref{fig:kappas}, where we choose a $3~\tev$ muon collider to illustrate this comparison in the parameter plane $\kappa_\mu$\nobreakdash-$\kappa_Z$. The shaded regions labeled by different values of the heavy Higgs mass 
show the higher signal rate from the RR process than both the $ZH$ associated production and $HA$ pair production. 
The nearly flat region for $1.4$ TeV $H$ and $A$ represents the good sensitivity from HA pair production in the low $\kappa_Z$ region. As expected, the RR process is more sensitive for a heavier Higgs boson near the energy threshold, which would be  especially important in the decoupling regime for $ZH/HA$ processes. 
At higher (lower) energies, the mass reach scales up (down), but with a lower (higher) luminosity need scaled by $1/s$. 

Only after specifying the underlying theory for the heavy Higgs bosons, and requiring the lighter Higgs
boson in agreement with the current LHC measurement, these two parameters could be constrained in a correlated manner, subject to the experimental accuracy. 
The allowed $\kappa_Z$ region is tightly constrained  by the currently observed SM-like Higgs boson. We reproduce the allowed parameter regions from Ref.~\cite{Barger:2013ofa}  for  four types of 2HDM with current LHC data (solid) and projection  after LHC-$300~\fbi$ (dashed). This illustrates that the RR processes is very much favored in 2HDM models, where the lighter SM-like Higgs boson carries most of the couplings to the electroweak gauge bosons.


\section{Direct Measurement of Higgs Invisible Branching Fraction}
\label{sec:invi}

The heavy Higgs boson could have deep connection with the dark matter sector, and have a sizable decay branching fraction to invisible particles~\cite{Birkedal:2006fz,MarchRussell:2008yu,Baer:2011ab,Han:2013gba,Han:2014nba}. We consider the signal of the Heavy Higgs thus rendered invisible in the context of the RR heavy Higgs production. The signal events contain a clean mono-chromatic photon that reconstructs the heavy Higgs mass without other particle activities.

\begin{figure}[t]
\centering
\includegraphics[scale=0.38]{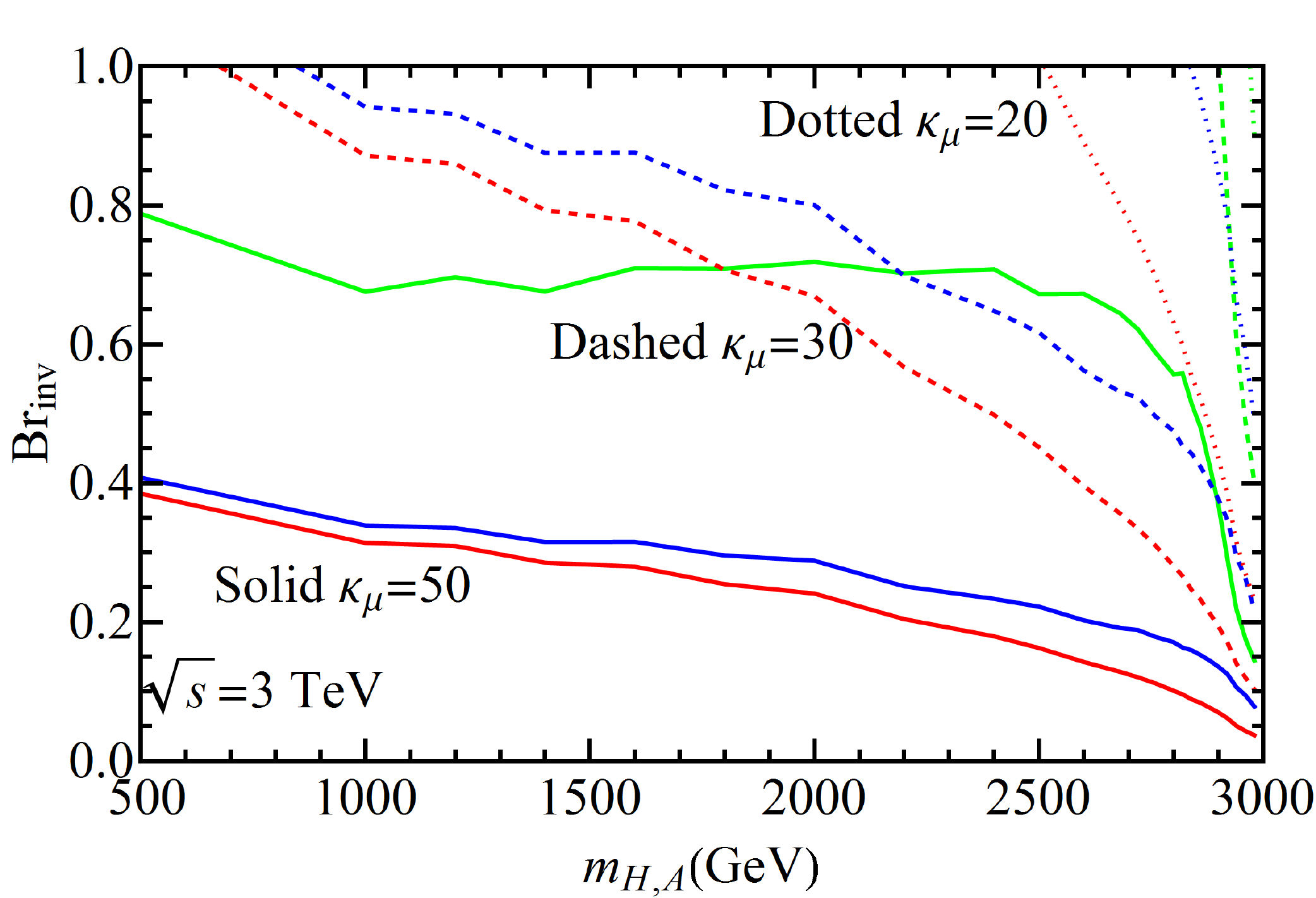}
\caption{Estimated $3\sigma$ reach for the invisible branching fraction of the heavy Higgs decay from RR at 3 TeV muon collider with 1.6 $\abi$. The Higgs widths are set at 1 (red), 10 (blue) and 100 (green) GeV.}
\label{fig:invbr}
\end{figure}

The $t$-channel $W$ boson exchange with ISR is the leading background ($\mm \to \nu\nu\gamma$), with its cross section as large as 2.6 pb. 
With this background included, we show the $3\sigma$ sensitivity to probe the invisible modes in Fig.~\ref{fig:invbr}. We exhibit the results for a series of $\kappa_\mu$ values (20, 30, 50) and Higgs widths (1, 10, 100 GeV) at the 3 TeV muon collider described in Table \ref{tab:MuC}. The choice of bin sizes is to optimize the signal background ratio for $\kappa_\mu=10$ at steps of 1 GeV.

We see that without knowing the heavy Higgs boson mass, one still gains some sensitivity for its invisible width. Once the mass is known from the RR process described in Sec.~\ref{sec:RR} or from other means, the invisible width can be probed by tuning the beam close to the resonance. The invisible and undetectable width can also be mapped out indirectly with a dedicated beam scan as well, similar to the muon collider Higgs factory~\cite{Han:2012rb,Conway:2013lca,Alexahin:2013ojp}.

\section{Conclusions}
\label{sec:sum}

We studied the signature and sensitivity for heavy Higgs boson signals from three production modes at a high energy muon collider. Compared to the $s$-channel resonance at $\sqrt s = m_h$, these different production mechanisms do not rely on {\it a priori} knowledge of the Higgs boson mass, and thus avoid the broad scanning procedure. We find that radiative return (RR) is of particular interest. This signal ($\gamma H$) is characterized by a mono-chromatic photon that yields a reconstructed recoil mass peak at the heavy Higgs boson mass.
We performed numerical simulations for this signal and its SM backgrounds and showed the coupling-mass parameter space $\kappa_{\mu}$\nobreakdash-$m$ (SUSY equivalent of $\tan\beta-M_{A}$) covered by such search at a high energy muon collider to be substantially extended over the LHC expectation with the direct observation of the heavy Higgs boson. 
Comparing with other modes of $ZH$ and $HA$ production at a lepton collider, the RR process is advantageous, especially for the ``decoupled'' scenarios in many 2HDM-like models. We further discussed its potential for measuring the invisible decays of the heavy Higgs boson and found some sensitivity especially for larger values of $\kappa_\mu$. The RR process could certainly provide us an interesting option comparing to traditional scanning procedure for heavy Higgs boson discovery at a high energy muon collider.

Because of the lepton universality for gauge interactions, the processes 
$\mu^+ \mu^- \to Z H,\ HA$ would be the same as those in $e^+ e^- $ collisions at the same c.m.~energy
since the contributions to both processes are overwhelmingly from the $s$-channel $Z$-exchange.
Thus the advantage of the RR process ($\gamma H$) would also apply when compared with a high energy $e^{+}e^{-}$ collider, where the RR process is essentially absent.

\begin{acknowledgments}
We would like to thank Estia Eichten for useful discussions.
T.H.~and Z.L.~are supported in part by the U.S.~Department of Energy under grant No. DE-FG02-95ER40896 and in part by PITT PACC. 
Z.L.~is supported in part by the Andrew Mellon Predoctoral Fellowship and a PITT PACC Predoctoral Fellowship  from Dietrich School of Art and Science, University of Pittsburgh. N.C.~and B.M.~acknowledge the funding available from the Department of Atomic Energy, Government of India, for the Regional Centre for Accelerator-based Particle Physics, Harish-Chandra Research Institute. T.H.~also thanks the Aspen Center for Physics for hospitality when this work was being finalized. ACP is supported by NSF under grant 1066293.

\end{acknowledgments}


\bibliographystyle{kp}
\bibliography{references}

\end{document}